# The Vertica Analytic Database: C-Store 7 Years Later

Andrew Lamb, Matt Fuller, Ramakrishna Varadarajan
Nga Tran, Ben Vandiver, Lyric Doshi, Chuck Bear
Vertica Systems, An HP Company
Cambridge, MA
{alamb, mfuller, rvaradarajan, ntran, bvandiver, ldoshi, cbear}
@vertica.com

## ABSTRACT

This paper describes the system architecture of the Vertica Analytic Database (Vertica), a commercialization of the design of the C-Store research prototype. Vertica demonstrates a modern commercial RDBMS system that presents a classical relational interface while at the same time achieving the high performance expected from modern "web scale" analytic systems by making appropriate architectural choices. Vertica is also an instructive lesson in how academic systems research can be directly commercialized into a successful product.

## 1. INTRODUCTION

The Vertica Analytic Database (Vertica) is a distributed[1], massively parallel RDBMS system that commercializes the ideas of the C-Store[21] project. It is one of the few new commercial relational database systems that is widely used in business critical systems. At the time of this writing, there are over 500 production deployments of Vertica, at least three of which are substantially over a petabyte in size. Despite the recent interest in academia and industry about so called "NoSQL" systems [13, 19, 12], the C-Store project anticipated the need for web scale distributed processing and these new NoSQL systems use many of the same techniques found in C-Store and other relational systems. Like any language or system, SQL is not perfect, but it has been a transformational abstraction for application developers, freeing them from many implementation details of storing and finding their data to focus their efforts on using the information effectively.

Vertica's experience in the marketplace and the emergence of other technologies such as Hive [7] and Tenzing [9] validate that the problem is **not** SQL. Rather, the unsuitability of legacy RDBMs systems for massive analytic workloads is that they were designed for transactional workloads on late-model computer hardware 40 years ago. Vertica is designed for analytic workloads on modern hardware and its success proves the commercial and technical viability of large scale distributed databases which offer fully ACID transactions yet efficiently process petabytes of structured data.

[1]We use the term *distributed database* to mean a shared-nothing, scale-out system as opposed to a set of locally autonomous (own catalog, security settings, etc.) RDBMS systems.

This main contributions of this paper are:

1. An overview of the architecture of the Vertica Analytic Database, focusing on deviations from C-Store.

2. Implementation and deployment lessons that led to those differences.

3. Observations on real-world experiences that can inform future research directions for large scale analytic systems.

We hope that this paper contributes a perspective on commercializing research projects and emphasizes the contributions of the database research community towards large scale distributed computing.

## 2. BACKGROUND

Vertica was the direct result of commercializing the C-Store research system. Vertica Systems was founded in 2005 by several of the C-Store authors and was acquired in 2011 by Hewlett-Packard (HP) after several years of commercial development. [2]. Significant research and development efforts continue on the Vertica Analytic Database.

### 2.1 Design Overview

#### 2.1.1 Design Goals

Vertica utilizes many (but not all) of the ideas of C-Store, but none of the code from the research prototype. Vertica was explicitly designed for analytic workloads rather than for transactional workloads.

**Transactional** workloads are characterized by a large number of transactions per second (e.g. thousands) where each transaction involves a handful of tuples. Most of the transactions take the form of single row insertions or modifications to existing rows. Examples are inserting a new sales record and updating a bank account balance.

**Analytic** workloads are characterized by smaller transaction volume (e.g. tens per second), but each transaction examines a significant fraction of the tuples in a table. Examples are aggregating sales data across time and geography dimensions and analyzing the behavior of distinct users on a web site.

As typical table sizes, even for small companies, have grown to millions and billions of rows, the difference between the transactional and analytic workloads has been increasing. As others have pointed out [26], it is possible to exceed the performance of existing one-size-fits-all systems by orders of magnitudes by focusing specifically on analytic workloads.

Vertica is a distributed system designed for modern commodity hardware. In 2012, this means x86_64 servers, Linux and commodity gigabit Ethernet interconnects. Like C-Store, Vertica is designed from the ground up to be a distributed database. When nodes are added to the database, the system's performance should scale linearly. To achieve such scaling, using a shared disk (often referred to as network-attached storage) is not acceptable as it almost immediately becomes a bottleneck. Also, the storage system's data placement, the optimizer and execution engine should avoid consuming large amounts of network bandwidth to prevent the interconnect from becoming the bottleneck.

In analytic workloads, while transactions per second is relatively low by Online-Transaction-Processing (OLTP) standards, rows processed per second is incredibly high. This applies not only to querying but also to loading the data into the database. Special care must be taken to support high ingest rates. If it takes days to load your data, a super-fast analytic query engine will be of limited use. Bulk load must be fast and must not prevent or unduly slow down queries ongoing in parallel.

For a real production system, all operations must be "online". Vertica can not require stopping or suspending queries for storage management or maintenance tasks. Vertica also aims to ease the management burden by making ease of use an explicit goal. We trade CPU cycles (which are cheap) for human wizard cycles (which are expensive) whenever possible. This takes many forms such as minimizing complex networking and disk setup, limiting performance tuning required, and automating physical design and management. All vendors claim management ease, though success in the real world is mixed.

Vertica was written entirely from scratch with the following exceptions, which are based on the PostgreSQL [5] implementation:

1. The SQL parser, semantic analyzer, and standard SQL rewrites.

2. Early versions of the standard client libraries, such as JDBC and ODBC and the command line interface.

All other components were custom written from the ground up. While this choice required significant engineering effort and delayed the initial introduction of Vertica to the market, it means Vertica is positioned to take full advantage of its architecture.

## 3. DATA MODEL

Like all SQL based systems, Vertica models user data as tables of columns (attributes), though the data is not physically arranged in this manner. Vertica supports the full range of standard INSERT, UPDATE, DELETE constructs for logically inserting and modifying data as well as a bulk loader and full SQL support for querying.

### 3.1 Projections

Like C-Store, Vertica physically organizes table data into *projections*, which are sorted subsets of the attributes of a table. Any number of projections with different sort orders and subsets of the table columns are allowed. Because Vertica is a column store and has been optimized so heavily for performance, it is **NOT** required to have one projection for each predicate that a user might restrict. In practice, most customers have one super projection (described below) and between zero and three narrow, non-super projections.

Each projection has a specific sort order on which the data is totally sorted as shown in Figure 1. Projections may be thought of as a restricted form of materialized view [11, 25]. They differ from standard materialized views because they are the only physical data structure in Vertica, rather than auxiliary indexes. Classical materialized views also contain aggregation, joins and other query constructs that Vertica projections do not. Experience has shown that the maintenance cost and additional implementation complexity of maintaining materialized views with aggregation and filtering is not practical in real world distributed systems. Vertica does support a special case to physically denormalize certain joins within prejoin projections as described below.

### 3.2 Join Indexes

C-Store uses a data structure called a *join index* to reconstitute tuples from the original table using different partial projections. While the authors expected only a few join indices in practice, Vertica does not implement join indices at all, instead requiring at least one *super projection* containing every column of the anchoring table. In practice and experiments with early prototypes, we found that the costs of using join indices far outweighed their benefits. Join indices were complex to implement and the runtime cost of reconstructing full tuples during distributed query execution was very high. In addition, explicitly storing row ids consumed significant disk space for large tables. The excellent compression achieved by our columnar design helped keep the cost of super projections to a minimum and we have no plans to lift the super projection requirement.

### 3.3 Prejoin Projections

Like C-Store, Vertica supports *prejoin* projections which permit joining the projection's anchor table with any number of dimension tables via N:1 joins. This permits a normalized logical schema, while allowing the physical storage to be denormalized. The cost of storing physically denormalized data is much less than in traditional systems because of the available encoding and compression. Prejoin projections are not used as often in practice as we expected. This is because Vertica's execution engine handles joins with small dimension tables very well (using highly optimized hash and merge join algorithms), so the benefits of a prejoin for query execution are not as significant as we initially predicted. In the case of joins involving a fact and a large dimension table or two large fact tables where the join cost is high, most customers are unwilling to slow down bulk loads to optimize such joins. In addition, joins during load offer fewer optimization opportunities than joins during query because the database knows nothing apriori about the data in the load stream.

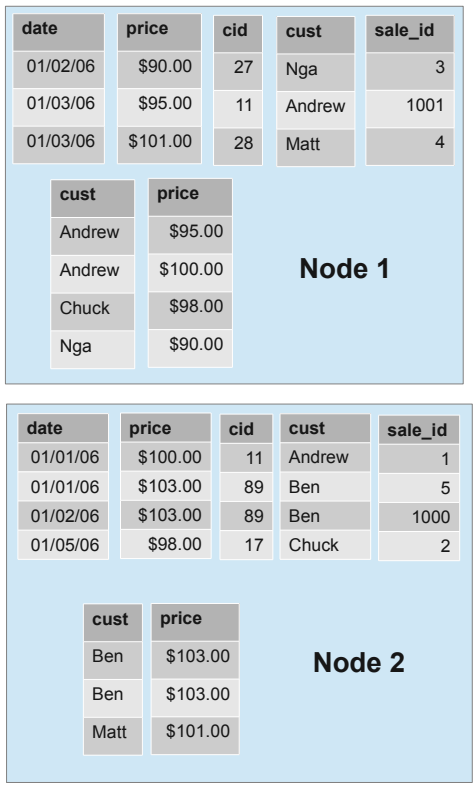

Figure 1: Relationship between tables and projections. The *sales* tables has 2 projections: (1) A super projection, sorted by date, segmented by $HASH(sale\_id)$ and (2) A non-super projection containing only $(cust, price)$ attributes, sorted by *cust*, segmented by $HASH(cust)$.

## 3.4 Encoding and Compression

Each column in each projection has a specific encoding scheme. Vertica implements a different set of encoding schemes than C-store, some of which are enumerated in Section 3.4.1. Different columns in a projection may have different encodings, and the same column may have a different encoding in each projection in which it appears. The same encoding schemes in Vertica are often far more effective than in other systems because of Vertica's sorted physical storage. A comparative illustration can be found in Section 8.2.

### 3.4.1 Encoding Types

1. **Auto**: The system automatically picks the most advantageous encoding type based on properties of the data itself. This type is the default and is used when insufficient usage examples are known.

2. **RLE**: Replaces sequences of identical values with a single pair that contains the value and number of occurrences. This type is best for low cardinality columns that are sorted.

3. **Delta Value**: Data is recorded as a difference from the smallest value in a data block. This type is best used for many-valued, unsorted integer or integer-based columns.

4. **Block Dictionary**: Within a data block, distinct column values are stored in a dictionary and actual values are replaced with references to the dictionary. This type is best for few-valued, unsorted columns such as stock prices.

5. **Compressed Delta Range**: Stores each value as a delta from the previous one. This type is ideal for many-valued float columns that are either sorted or confined to a range.

6. **Compressed Common Delta**: Builds a dictionary of all the deltas in the block and then stores indexes into the dictionary using entropy coding. This type is best for sorted data with predictable sequences and occasional sequence breaks. For example, timestamps recorded at periodic intervals or primary keys.

## 3.5 Partitioning

C-Store mentions intra-node "Horizontal Partitioning" as a way to improve performance by increasing parallelism within a single node. In contrast, Vertica's execution engine, as described in Section 6.1, obtains intra-node parallelism without requiring separation of the on-disk physical structures. It does so by dividing each on-disk structure into logical regions at runtime and processing the regions in parallel. Despite the automatic parallelization, Vertica does provide a way to keep data segregated in physical structures based on value through a simple syntax:
`CREATE TABLE ... PARTITION BY <expr>`.
This instructs Vertica to maintain physical storage so that all tuples within a ROS container[2] evaluate to the same distinct value of the partition expression. Partition expressions are most often date related such as extracting the month and year from a timestamp.

---
[2]ROS and ROS containers are explained in section 3.7

1792

The first reason for partitioning, as in other RDBMS systems, is fast bulk deletion. It is common to keep data separated into files based on a combination of month and year, so removing a specific month of data from the system is as simple as deleting files from a filesystem. This arrangement is very fast and reclaims storage immediately. The alternative, if the data is not pre-separated, requires searching all physical files for rows matching the delete predicate and adding delete vectors (further explained in Section 3.7.1) for deleted records. It is much slower to find and mark deleted records than deleting files, and this procedure actually increases storage requirements and degrades query performance until the tuple mover's next merge-out operation is performed (see Section 4). Because bulk deletion is only fast if all projections are partitioned the same way, partitioning is specified at the table level and not the projection level.

The second way Vertica takes advantage of physical storage separation is improving query performance. As described here [22], Vertica stores the minimum and maximum values of the column data in each ROS to quickly prune containers at plan time that can not possibly pass query predicates. Partitioning makes this technique more effective by preventing intermixed column values in the same ROS.

## 3.6 Segmentation: Cluster Distribution

C-Store separates physical storage into segments based on the first column in the sort order of a projection and the authors briefly mention their plan to design a storage allocator for assigning segments to nodes. Vertica has a fully implemented distributed storage system that assigns tuples to specific computation nodes. We call this internode (splitting tuples among nodes) horizontal partitioning *segmentation* to distinguish it from the intra-node (segregating tuples within nodes) partitioning described in Section 3.5. Segmentation is specified for each projection, which can be (and most often is) different from the sort order. Projection segmentation provides a deterministic mapping of tuple value to node and thus enables many important optimizations. For example, Vertica uses segmentation to perform fully local distributed joins and efficient distributed aggregations, which is particularly effective for the computation of high-cardinality distinct aggregates

Projections can either be *replicated* or *segmented* on some or all cluster nodes. As the name implies, a replicated projection stores a copy of each tuple on every projection node. Segmented projections store each tuple on exactly one specific projection node. The node on which the tuple is stored is determined by a segmentation clause in the projection definition: `CREATE PROJECTION ... SEGMENTED BY <expr>` where `<expr>` is an arbitrary [3] integral expression.

Nodes are assigned to store ranges of segmentation expression values, starting with the following mapping where $C_{MAX}$ is the maximum integral value ($2^{64}$ in Vertica).

$$
\begin{array}{lll}
0 \leq expr < \frac{C_{MAX}}{N} & \Rightarrow Node_1 \\
\frac{1*C_{MAX}}{N} \leq expr < \frac{2*C_{MAX}}{N} & \Rightarrow Node_2 \\
\quad \ldots & \ldots \\
\frac{(N-2)*C_{MAX}}{N} \leq expr < \frac{(N-1)*C_{MAX}}{N} & \Rightarrow Node_{N-1} \\
\frac{(N-1)*C_{MAX}}{N} \leq expr < C_{MAX} & \Rightarrow Node_N
\end{array}
$$

---
[3]While it is possible to manually specify segmentation, most users let the Database Designer determine an appropriate segmentation expression for projections.

This is a classic ring style segmentation scheme. The most common choice is $HASH(col_1..col_n)$, where $col_i$ is some suitably high cardinality column with relatively even value distributions, commonly a primary key column. Within each node, in addition to the specified partitioning, Vertica keeps tuples physically segregated into "local segments" to facilitate online expansion and contractions of the cluster. When nodes are added or removed, data is quickly transferred by assigning one or more of the existing local segments to a new node and transferring the segment data wholesale in its native format, without any rearrangement or splitting necessary.

## 3.7 Read and Write Optimized Stores

Like C-Store, Vertica has a Read Optimized Store (ROS) and a Write Optimized Store (WOS). Data in the ROS is physically stored in multiple *ROS containers* on a standard file system. Each ROS container logically contains some number of complete tuples sorted by the projection's sort order, stored as a pair of files per column. Vertica is a true column store – column files may be independently retrieved as the storage is physically separate. Vertica stores two files per column within a ROS container: one with the actual column data, and one with a *position index*. Data is identified within each ROS container by a *position* which is simply its ordinal position within the file. Positions are implicit and are never stored explicitly. The position index is approximately $\frac{1}{1000}$ the size of the raw column data and stores metadata per disk block such as start position, minimum value and maximum value that improve the speed of the execution engine and permits fast tuple reconstruction. Unlike C-Store, this index structure does not utilize a B-Tree as the ROS containers are never modified. Complete tuples are reconstructed by fetching values with the same position from each column file within a ROS container. Vertica also supports grouping multiple columns together into the same file when writing to a ROS container. This hybrid row-column storage mode is very rarely used in practice because of the performance and compression penalty it exacts.

Data in the WOS is solely in memory, where column or row orientation doesn't matter. The WOS's primary purpose is to buffer small data inserts, deletes and updates so that writes to physical structures contain a sufficient numbers of rows to amortize the cost of the writing. The WOS has changed over time from row orientation to column orientation and back again. We did not find any significant performance differences between these approaches and the changes were driven primarily by software engineering considerations. Data is not encoded or compressed when it is in the WOS. However, it *is* segmented according to the projection's segmentation expression.

### 3.7.1 Data Modifications and Delete Vectors

Data in Vertica is never modified in place. When a tuple is deleted or updated from either the WOS or ROS, Vertica creates a *delete vector*. A delete vector is a list of positions of rows that have been deleted. Delete vectors are stored in the same format as user data: they are first written to a *DVWOS* in memory, then moved to *DVROS containers* on disk by the tuple mover (further explained in section 4) and stored using efficient compression mechanisms. There may be multiple delete vectors for the WOS and multiple delete vectors for any particular ROS container. SQL `UPDATE` is supported by deleting the row being updated and then inserting a row containing the updated column values.

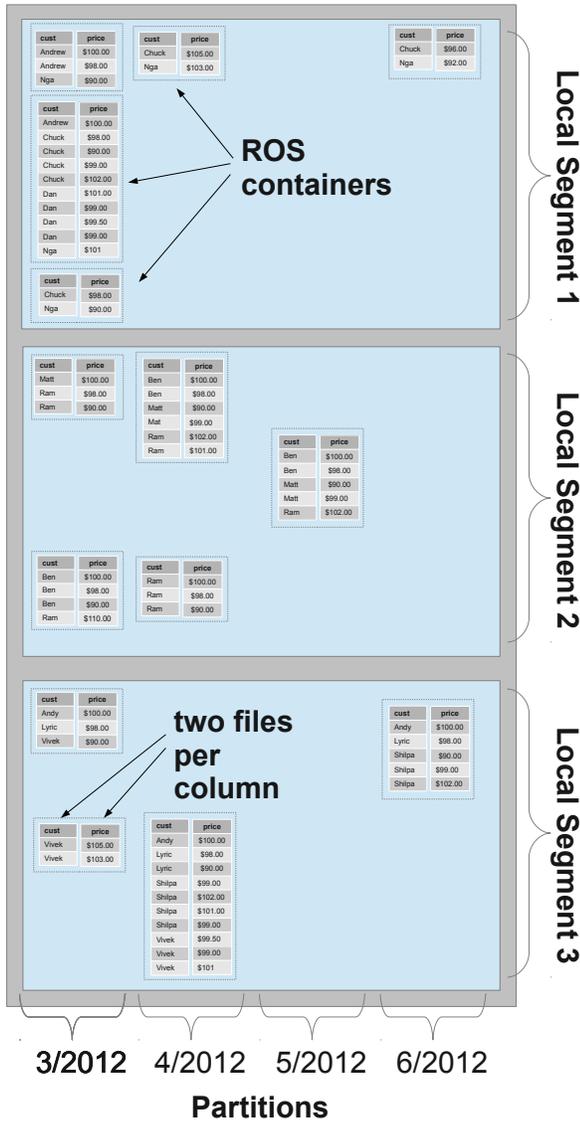

**Figure 2: Physical storage layout within a node.**
This figure illustrates how columns are stored in projections using files on disk. The table is partitioned by EXTRACT MONTH, YEAR FROM TIMESTAMP and segmented by HASH(cid). There are 14 ROS containers, each with two columns. Each column's data within its ROS container is stored as a single file for a total of 28 files of user data. The data has four partition keys: 3/2012, 4/2012, 5/2012 and 6/2012. As the projection is segmented by HASH(cid), this node is responsible for storing all data that satisfies the condition $C_{nmin} < hash(cid) \leq C_{nmax}$, for some value of $C_{nmin}$ and $C_{nmax}$. This node has divided the data into three local segments such that: Local Segment 1 has $C_{nmin} < hash(cid) \leq \frac{C_{nmax}}{3}$, Local Segment 2 has $\frac{C_{nmin}}{3} < hash(cid) \leq \frac{2*C_{nmax}}{3}$ and Local Segment 3 has $\frac{2*C_{nmin}}{3} < hash(cid) \leq C_{nmax}$.

## 4. TUPLE MOVER

The *tuple mover* is an automatic system which oversees and rearranges the physical data files to increase data storage and ingest efficiency during query processing. Its work can be grouped into two main functions:

1. **Moveout**: asynchronously moves data from the WOS to the ROS
2. **Mergeout**: merges multiple ROS files together into larger ones.

As the WOS fills up, the tuple mover automatically executes a *moveout* operation to move data from WOS to ROS. In the event that the WOS becomes saturated before moveout is complete, subsequently loaded data is written directly to new ROS Containers until the WOS regains sufficient capacity. The tuple mover must balance its moveout work so that it is not overzealous (creating too many little ROS containers) but also not too lazy (resulting in WOS overflow which also creates too many little files).

*Mergeout* decreases the number of ROS containers on disk. Numerous small ROS containers decrease compression opportunities and slow query processing. Many files require more file handles, more seeks, and more merges of the sorted files. The tuple mover merges smaller files together into larger ones, and it reclaims storage by filtering out tuples which were deleted prior to the Ancient History Mark (further explained in section 5.1) as there is no way a user can query them. Unlike C-Store, the tuple mover does not intermix data from the WOS and ROS in order to strongly bound the number of times a tuple is (re)merged. When a tuple is part of a mergeout operation, it is read from disk once and written to disk once.

The tuple mover periodically quantizes the ROS containers into several exponential sized *strata* based on file size. The output ROS container from a mergeout operation are planned such that the resulting ROS container is in at least one strata larger than any of the input ROS containers. Vertica does not impose any size restrictions on ROS containers, but the tuple mover will not create ROS containers greater than some maximum (currently 2TB) so as to strongly bound the number of strata and thus the number of merges. The maximum ROS container size is chosen to be sufficiently large that any per-file overhead is amortized to irrelevance and yet the files are not too unwieldy to manage. By choosing strata sizes exponentially, the number of times any tuple is rewritten is bounded to the number of strata.

The tuple mover takes care to preserve partition and local segment boundaries when choosing merge candidates. It has also been tuned to maximize the system's tuple ingest rate while preventing an explosion in the number of ROS containers. An important design point of the tuple mover is that operations are not centrally coordinated across the cluster. The specific ROS container layouts are private to every node, and while two nodes might contain the same tuples, it is common for them to have a different layout of ROS containers due to factors such as different patterns of merging, available resources, node failure and recovery.

| Requested Mode | Granted Mode ||||||| 
| --- | --- | --- | --- | --- | --- | --- | --- |
| | S | I | SI | X | T | U | O |
| S | Yes | No | No | No | Yes | Yes | No |
| I | No | Yes | No | No | Yes | Yes | No |
| SI | No | No | No | No | Yes | Yes | No |
| X | No | No | No | No | No | Yes | No |
| T | Yes | Yes | Yes | No | Yes | Yes | No |
| U | Yes | Yes | Yes | Yes | Yes | Yes | No |
| O | No | No | No | No | No | No | No |

Table 1: Lock Compatibility Matrix

| Requested Mode | Granted Mode ||||||| 
| --- | --- | --- | --- | --- | --- | --- | --- |
| | S | I | SI | X | T | U | O |
| S | S | SI | SI | X | S | S | O |
| I | SI | I | SI | X | I | I | O |
| SI | SI | SI | SI | X | SI | SI | O |
| X | X | X | X | X | X | X | O |
| T | S | I | SI | X | T | T | O |
| U | S | I | SI | X | T | U | O |
| O | O | O | O | O | O | O | O |

Table 2: Lock Conversion Matrix

## 5. UPDATES AND TRANSACTIONS

Every tuple in Vertica is timestamped with the logical time at which it was committed. Each delete marker is paired with the logical time the row was deleted. These logical timestamps are called epochs and are implemented as implicit 64-bit integral columns on the projection or delete vector. All nodes agree on the epoch in which each transaction commits, thus an epoch boundary represents a globally consistent snapshot. In concert with Vertica's policy of never modifying storage, a query executing in the recent past needs no locks and is assured of a consistent snapshot. The default transaction isolation in Vertica is READ COMMITTED, where each query targets the latest epoch (the current epoch - 1).

Because most queries, as explained above, do not require any locks, Vertica has an analytic-workload appropriate table locking model. Lock compatibility and conversion matrices are shown in Table 1 and Table 2 respectively, both adapted from [15].

- **S**hared lock: while held, prevents concurrent modification of the table. Used to implement SERIALIZABLE isolation.

- **I**nsert lock: required to insert data into a table. An **I**nsert lock is compatible with itself, enabling multiple inserts and bulk loads to occur simultaneously which is critical to maintain high ingest rates and parallel loads yet still offer transactional semantics.

- **S**hared**I**nsert lock: required for read and insert, but not update or delete.

- E**X**clusive lock: required for deletes and updates.

- **T**uple mover lock: required for certain tuple mover operations. This lock is compatible with every lock except **X** and is used by the tuple mover during certain short operations on delete vectors.

- **U**sage lock: required for parts of moveout and mergeout operations.

- **O**wner lock: required for significant DDL such as dropping partitions and adding columns.

Vertica employs a distributed agreement and group membership protocol to coordinate actions between nodes in the cluster. The messaging protocol uses broadcast and point-to-point delivery to ensure that any control message is successfully received by every node. Failure to receive a message will cause a node to be ejected from the cluster and the remaining nodes will be informed of the loss. Like C-Store, Vertica does not employ traditional two-phase commit[15]. Rather, once a cluster transaction commit message is sent, nodes either successfully complete the commit or are ejected from the cluster. A commit succeeds on the cluster if it succeeds on a quorum of nodes. Any ROS or WOS created by the committing transaction becomes visible to other transactions when the commit completes. Nodes that fail during the commit process leave the cluster and rejoin the cluster in a consistent state via the recovery mechanism described in section 5.2. Transaction rollback simply entails discarding any ROS container or WOS data created by the transaction.

### 5.1 Epoch Management

Originally, Vertica followed the C-store epoch model: epochs contained all transactions committed in a given time window. However, users running in READ COMMITTED were often confused because their commits did not become "visible" until the epoch advanced. Now Vertica automatically advances the epoch as part of commit when the committing transaction includes DML or certain data-modifying DDL. In addition to reducing user confusion, automatic epoch advancement simplifies many of the internal management processes (like the tuple mover).

Vertica tracks two epoch values worthy of mention: the *Last Good Epoch* (LGE) and the *Ancient History Mark* (AHM). The LGE for a node is the epoch for which all data has been successfully moved out of the WOS and into ROS containers on disk. The LGE is tracked per projection because data that exists only in the WOS is lost in the event of a node failure. The AHM is an analogue of C-store's low water mark where Vertica discards historical information prior to the AHM when data reorganization occurs. Whenever the tuple mover observes a row deleted prior to the AHM, it elides the row from the output of the operation. The AHM advances automatically according to a user-specified policy. The AHM normally does not advance when nodes are down so as to preserve the history necessary to incrementally replay DML operations during recovery (described in Section 5.2).

### 5.2 Tolerating Failures

Vertica replicates data to provide fault tolerance by employing the projection segmentation mechanism explained in section 3.6. Each projection must have at least one *buddy projection* containing the same columns and a segmentation method that ensures that no row is stored on the same node by both projections. When a node is down, the buddy projection is employed to source the missing rows from the down node. Like any distributed database, Vertica must grace-

1795

fully handle failed nodes rejoining the cluster. In Vertica, this process is called *recovery*. Vertica has no need of traditional transaction logs because the data+epoch itself serves as a log of past system activity. Vertica implements efficient incremental recovery by utilizing this historical record to replay DML the down node has missed. When a node rejoins the cluster after a failure, it recovers each projection segment from a corresponding buddy projection segment. First, the node truncates all tuples that were inserted after its LGE, ensuring that it starts at a consistent state. Then recovery proceeds in two phases to minimize operational disruption.

- **Historical Phase**: recovers committed data from the LGE to some previous epoch $E_h$. No locks are held while data between the recovering node's LGE and $E_h$ is copied from the buddy projection. When complete, the projection's LGE is advanced to $E_h$ and either the historical phase continues or the current phase is entered, depending on the amount of data between the new LGE and the current epoch.

- **Current Phase**: recovers committed data from the LGE until the current epoch. The current phase takes a **S**hared lock on the projection's tables and copies any remaining data. After the current phase, recovery is complete and the projection participates in all future DML transactions.

If the projection and its buddy have matching sort orders, recovery simply copies whole ROS containers and their delete vectors from one node to another. Otherwise, an execution plan similar to `INSERT ... SELECT ...` is used to move rows (including deleted rows) to the recovering node. A separate plan is used to move delete vectors. The *refresh* and *rebalance* operations are similar to the recovery mechanism. Refresh is used to populate new projections which were created after the table was loaded with data. Rebalance redistributes segments between nodes to rebalance storage as nodes are added and removed. Both have a historical phase where older data is copied and a current phase where a **S**hared lock is held while any remaining data is transferred.

Backup is handled completely differently by taking advantage of Vertica's read-only storage system. A backup operation takes a snapshot of the database catalog and creates hard-links for each Vertica data file on the file system. The hard-links ensure that the data files are not removed while the backup image is copied off the cluster to the backup location. Afterwards, the hard-links are removed, ensuring that storage used by any files artificially preserved by the backup is reclaimed. The backup mechanism supports both full and incremental backup.

Recovery, refresh, rebalance and backup are all online operations; Vertica continues to load and query data while they are running. They impact ongoing operations only to the extent that they require computational and bandwidth resources to complete.

## 5.3 Cluster Integrity

The primary state managed between the nodes is the metadata catalog, which records information about tables, users, nodes, epochs, etc. Unlike other databases, the catalog is not stored in database tables, as Vertica's table design is inappropriate for catalog access and update. Instead, the catalog is implemented using a custom memory resident data structure and transactionally persisted to disk via its own mechanism, both of which are beyond the scope of this paper.

As in C-Store, Vertica provides the notion of *K-safety*: With $K$ or fewer nodes down, the cluster is guaranteed to remain available. To achieve K-Safety, the database projection design must ensure at least $K+1$ copies of each segment are present on different nodes such that a failure of **any** $K$ nodes leaves at least one copy available. The failure of $K+1$ nodes does not guarantee a database shutdown. Only when node failures actually cause data to become unavailable will the database shutdown until the failures can be repaired and consistency restored via recovery. A Vertica cluster will also perform a safety shutdown if $\frac{N}{2}$ nodes are lost where $N$ is the number of nodes in the cluster. The agreement protocol requires a $\frac{N}{2}+1$ quorum to protect against network partitions and avoid a split brain effect where two halves of the cluster continue to operate independently.

## 6. QUERY EXECUTION

Vertica supports the standard SQL declarative query language along with its own proprietary extensions. Vertica's extensions are designed for cases where easily querying time-series and log style data in SQL was overly cumbersome or impossible. Users submit SQL queries using an interactive `vsql` command prompt or via standard JDBC, ODBC, or ADO .net drivers. Rather than continuing to add more proprietary extensions, Vertica has chosen to add an SDK with hooks for users to extend various parts of the execution engine.

### 6.1 Query Operators and Plan Format

The data processing of the plan is performed by the Vertica *Execution Engine* (EE). A Vertica query plan is a standard tree of operators where each operator is responsible for performing a certain algorithm. The output of one operator serves as the input to the following operator. A simple single node plan is illustrated in figure 3. Vertica's execution engine is multi-threaded and pipelined: more than one operator can be running at any time and more than one thread can be executing the code for any individual operator. As in C-store, the EE is fully vectorized and makes requests for blocks of rows at a time instead of requesting single rows at a time. Vertica's operators use a pull processing model: the most downstream operator requests rows from the next operator upstream in the processing pipeline. This operator does the same until a request is made of an operator that reads data from disk or the network. The available operator types in the EE are enumerated below. Each operator can use one of several possible algorithms which are automatically chosen by the query optimizer.

1. **Scan**: Reads data from a particular projection's ROS containers, and applies predicates in the most advantageous manner possible.

2. **GroupBy**: Groups and aggregates data. We have several different hash based algorithms depending on what is needed for maximal performance, how much memory is allotted, and if the operator must produce unique groups. Vertica also implements classic pipelined (one-pass) aggregates, with a choice to keep the incoming data encoded or not.

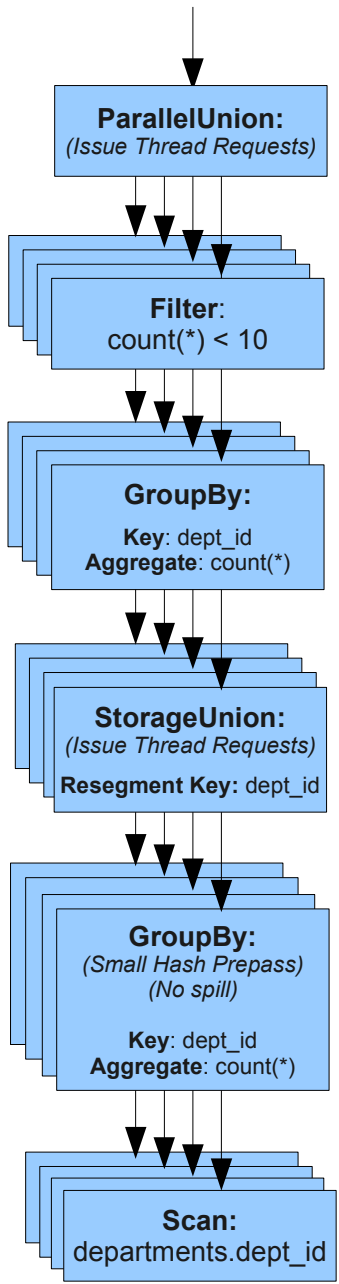

```
select dept_id, count(*)
from departments
group by dept_id
having count(*) < 10;
```

Figure 3: Plan representing a SQL query. The query plan contains a scan operator for reading data followed by operators for grouping and aggregation finally followed by a filter operation. The StorageUnion dispatches threads for processing data on a set of ROS containers. The StorageUnion also locally resegments the data for the above GroupBys. The ParallelUnion dispatches threads for processing the GroupBys And Filters in parallel.

3. **Join**: Performs classic relational join. Vertica supports both hash join and merge join algorithms which are capable of externalizing if necessary. All flavors of INNER, LEFT OUTER, RIGHT OUTER, FULL OUTER, SEMI, and ANTI joins are supported.

4. **ExprEval**: Evaluate an expression

5. **Sort**: Sorts incoming data, externalizing if needed.

6. **Analytic**: Computes SQL-99 Analytics style windowed aggregates

7. **Send/Recv**: Sends tuples from one node to another. Both broadcast and sending to nodes based on segmentation expression evaluation is supported. Each Send and Recv pair is capable of retaining the sortedness of the input stream.

Vertica's operators are optimized for the sorted data that the storage system maintains. Like C-Store, significant care has been taken and implementation complexity has been added to ensure operators can operate directly on encoded data, which is especially important for scans, joins and certain low level aggregates.

The EE has several techniques to achieve high performance. Sideways Information Passing (SIP) has been effective in improving join performance by filtering data as early as possible in the plan. It can be thought of as an advanced variation of predicate push down since the join is being used to do filtering [27]. For example, consider a HashJoin that joins two tables using simple equality predicates. The HashJoin will first create a hash table from the inner input before it starts reading data from the outer input to do the join. Special SIP filters are built during optimizer planning and placed in the Scan operator. At run time, the Scan has access to the Join's hash table and the SIP filters are used to evaluate whether the outer key values exist in the hash table. Rows that do not pass these filters are not output by the Scan thus increaseing performance since we are not unnecessarily bringing the data through the plan only to be filtered away later by the join. Depending on the join type, we are not always able to push the SIP filter to the Scan, but we do push the filters down as far as possible. We can also perform SIP for merge joins with a slightly different type of SIP filter beyond the scope of this paper.

The EE also switches algorithms during runtime as it observes data flowing through the system. For example, if Vertica determines at runtime the hash table for a hash join will not fit into memory, we will perform a sort-merge join instead. We also institute several "prepass" operators to compute partial results in parallel but which are not required for correctness (see Figure 3). The results of prepass operators are fed into the final operator to compute the complete result. For example, the query optimizer plans grouping operations in several stages for maximal performance. In the first stage, it attempts to aggregate immediately after fetching columns off the disk using an L1 cache sized hash table. When the hash table fills up, the operator outputs its current contents, clears the hash table, and starts aggregating afresh with the next input. The idea is to cheaply reduce the amount of data before sending it through other operators in the pipeline. Since there is still a small, but non-zero cost to run the prepass operator, the EE will decide at runtime to stop if it is not actually reducing the number of rows which pass.

During query compile time, each operator is given a memory budget based on the resources available given a user defined workload policy and what each operator is going to do. All operators are capable of handling arbitrary sized inputs, regardless of the memory allocated, by externalizing their buffers to disk. This is critical for a production database to ensure users queries are always answered. One challenge of a fully pipelined execution engine such as Vertica's is that all operators must share common resources, potentially causing unnecessary spills to disk. In Vertica, the plan is separated into multiple zones that can not be executing at the same time[4]. Downstream operators are able to reclaim resources previously used by upstream operators, allowing each operator more memory than if we pessimistically to assumed all operators would need their resources at the same time.

Many computations are data type dependent and require the code to branch to type specific implementations at query runtime. To improve performance and reduce control flow overhead, Vertica uses just in time compilation of certain expression evaluations to avoid branching by compiling the necessary assembly code on the fly.

Although the simplest implementation of a pull execution engine is a single thread, Vertica uses multiple threads for processing the same plan. For example, multiple worker threads are dispatched to fetch data from disk and perform initial aggregations on non overlapping sections of ROS containers. The Optimizer and EE work together to combine the data from each pipeline at the required locations to get correct answers. It is necessary to combine partial results because alike values are not co-located in the same pipeline.

The Send and Recv operators ship data to the nodes in the cluster. The send operator is capable of segmenting the data in such as way that all alike values are sent to the same node in the cluster. This allows each node's operator to compute the full results independently of the other nodes. In the same way we fully utilize the cluster of nodes by dividing the data in advantageous ways, we can divide the data locally on each node to process data in parallel and keep the all the cores fully utilized. As shown in Figure 3, multiple GroupBy operators are run in parallel requesting data from the StorageUnion which resegments the data such that the GroupBy is able to compute complete results.

## 6.2 Query Optimization

C-Store has a minimal optimizer, in which the projections it reaches first are chosen for tables in the query, and the join order of the projections is completely random. The Vertica Optimizer has evolved through three generations: StarOpt, StarifiedOpt, and V2Opt.

*StarOpt*, the initial Vertica optimizer, was a Kimball-style optimizer[18] which assumed that any interesting warehouse schema could be modeled as a classic star or snowflake. A star schema classifies attributes of an event into fact tables and descriptive attributes into dimension tables. Usually, a fact table is much larger than a dimension table and has a many-to-one relationship with its associated descriptive dimension tables. A snowflake schema is an extension of a star schema, where one or more dimension tables has many-to-one relationships with further descriptive dimension tables. This paper uses the term `star` to represent both star and snowflake designs. The information of a star schema is often requested through (star) queries that join fact tables with their dimensions. Efficient plans for star queries are to join a fact table with its most highly selective dimensions first. Thus, the most important process in planning a Vertica StarOpt query is choosing and joining projections with highly compressed and sorted predicate and join columns, to make sure that not only fast scans and merge joins on compressed columns are applied first, but also that the cardinality of the data for later joins is reduced.

Besides the described StarOpt and columnar-specific techniques described above, StartOpt and two other Vertica optimizers described later also employ other techniques to take advantage of the specifics of sorted columnar storage and compression, such as late materialization[8], compression-aware costing and planning, stream aggregation, sort elimination, and merge joins.

Although Vertica has been a distributed system since the beginning, StarOpt was designed only to handle queries whose tables have data with co-located projections. In other words, projections of different tables in the query must be either replicated on all nodes, or segmented on the same range of data on their join keys, so the plan can be executed locally on each node and the results sent to to the node that the client is connected to. Even with this limitation, StarOpt still works well with star schemas because only the data of large fact tables needs to be segmented throughout the cluster. Data of small dimension tables can be replicated everywhere without performance degradation. As many Vertica customers demonstrated their increasing need for non-star queries, Vertica developed its second generation optimizer, *StarfiedOpt*[5] as a modication to StarOpt. By forcing non-star queries to look like a star, Vertica could run the StarOpt algorithm on the query to optimize it. StarifiedOpt is far more effective for non-star queries than we could have reasonably hoped, but, more importantely, it bridged the gap to optimize both star and non-star queries while we designed and implemented the third generation optimizer: the custom built V2Opt.

The distribution aware *V2Opt*[6], which allows data to be transferred on-the-fly between nodes of the cluster during query execution, is designed from the start as a set of extensible modules. In this way, the brains of the optimizer can be changed without rewriting lots of the code. In fact, due to the inherent extensible design, knowledge gleaned from end-user experiences has already been incorporated into the V2Opt optimizer without a lot of additional engineering effort. V2Opt plans a query by categorizing and classifying the query's physical-properties, such as column selectivity, projection column sort order, projection data segmentation, prejoin projection availability, and integrity constraint availability. These physical-property heuristics, combined with a pruning strategy using a cost-model, based on compression aware I/O, CPU and Network transfer costs, help the optimizer (1) control the explosion in search space while continuing to explore optimal plans and (2) account for data distribution and bushy plans during the join order enumeration phase. While innovating on the V2Opt core algorithms, we also incorporated many of the best practices developed over the past 30 years of optimizer research such as using equi-height histograms to calculate selectivity, applying sample-based estimates of the number of distinct values [16], introducing transitive predicates based on join keys, converting outer joins to inner joins, subquery de-correlation, subquery flattening [17], view flattening, optimizing queries to favor co-located joins where possible, and automatically pruning out unnecessary parts of the query.

---

[4]Separated by operators such as Sort

[5]US patent 8,086,598, Query Optimizer with Schema Conversion

[6]Pending patent, Modular Query Optimizer

The Vertica Database Designer described in Section 6.3 works hand-in-glove with the optimizer to produce a physical design that takes advantage of the numerous optimization techniques available to the optimizer. Furthermore, when one or more nodes in the database cluster goes down, the optimizer replans the query by replacing and then re-costing the projections on unavailable nodes with their corresponding buddy projections on working nodes. This can lead to a new plan with a different join order from the original one.

## 6.3 Automatic Physical Design

Vertica features an automatic physical design tool called the *Database Designer* (DBD). The physical design problem in Vertica is to determine sets of projections that optimize a representative query workload for a given schema and sample data while remaining within a certain space budget. The major tensions to resolve during projection design are optimizing query performance while reducing data load overhead and minimizing storage footprint.

The DBD design algorithm has two sequential phases:

1. **Query Optimization**: Chooses projection sort order and segmentation to optimize the performance of the query workload. During this phase, the DBD enumerates candidate projections based on heuristics such as predicates, group by columns, order by columns, aggregate columns, and join predicates. The optimizer is invoked for each input query and given a choice of the candidate projections. The resulting plan is used to choose the best projections from amongst the candidates. The DBD's system to resolve conflicts when different queries are optimized by different projections is important, but beyond the scope of this paper. The DBD's direct use of the optimizer and cost model guarantees that it remains synchronized as the optimizer evolves over time.

2. **Storage Optimization**: Finds the best encoding schemes for the designed projections via a series of empirical encoding experiments on the sample data, given the sort orders chosen in the query optimization phase.

The DBD provides different design policies so users can trade off query optimization and storage footprint: (a) load-optimized, (b) query-optimized and (c) balanced. These policies indirectly control the number of projections proposed to achieve the desired balance between query performance and storage/load constraints. Other design challenges include monitoring changes in query workload, schema, and cluster layout and determining the incremental impact on the design.

As our user base has expanded, the DBD is now universally used for a baseline physical design. Users can then manually modify the proposed design before deployment. Especially in the case of the largest (and thus most important) tables, expert users sometimes make minor changes to projection-segmentation, select-list or sort-list based on their specific knowledge of their data or use cases which may be unavailable to the DBD. It is extremely rare for any user to override the column encoding choices of the DBD, which we credit to the empirical measurement during the storage-optimization phase.

## 7. USER EXPERIENCE

In this section we highlight some of the features of our system which have led to its wide adoption and commercial success, as well as the observations which led us to those features.

- **SQL**: First and foremost, standard SQL support was critical for commercial success as most customer organizations have large skill and tool investments in the language. Despite the temptation to invent new languages or dialects to avoid pet peeves, [7] standard SQL provides a data management system of much greater reach than a new language that people must learn.

- **Resource Management**: Specifying how a cluster's resources are to be shared and reporting on the current resource allocation with many concurrent users is critical to real world deployments. We initially under appreciated this point early in Vertica's lifetime and we believe it is still an understudied problem in academic data management research.

- **Automated Tuning**: Database users by and large wish to remain ignorant of a database's inner workings and focus on their application logic. Legacy RDBMS systems often require heroic tuning efforts, which Vertica has largely avoided by significant engineering effort and focus. For example, performance of early beta versions was a function of the physical storage layout and required users to learn how to tune and control the storage system. Automating storage layout management required Vertica to make significant and interrelated changes to the storage system, execution engine and tuple mover.

- **Predictability vs. Special Case Optimizations**: It was tempting to pick low hanging performance optimization fruit that could be delivered quickly, such as transitive predicate creation for INNER but not OUTER joins or specialized filter predicates for Hash joins but not Merge joins. To our surprise, such special case optimizations caused almost as many problems as they solved because certain user queries would go super fast and some would not in hard to predict ways, often due to some incredibly low level implementation detail. To our surprise, users didn't accept the rationale that it was better that some queries got faster even though not all did.

- **Direct Loading to the ROS**: While appealing in theory, directing all newly-inserted data to the WOS wastefully consumes memory. Especially while initially loading a system, the amount of data in a single bulk load operation was likely to be many tens of gigabytes in size and thus not memory resident. Users are more than happy to explicitly tag such loads to target the ROS in exchange for improved resource usage.

---
[7]Which at least one author admits having done in the past

| Metric | C-Store | Vertica |
|---|---|---|
| Q1 | 30 ms | 14 ms |
| Q2 | 360 ms | 71 ms |
| Q3 | 4900 ms | 4833 ms |
| Q4 | 2090 ms | 280 ms |
| Q5 | 310 ms | 93 ms |
| Q6 | 8500 ms | 4143 ms |
| Q7 | 2540 ms | 161 ms |
| Total Query Time | 18.7 s | 9.6s |
| Disk Space Required | 1,987 MB | 949 MB |

Table 3: Performance of Vertica compared with C-Store on single node Pentium 4 hardware using the queries and test harness of the C-Store paper.

| | Size (MB) | Comp. Ratio | Bytes Per Row |
|---|---|---|---|
| **Rand. Integers** | | | |
| Raw | 7.5 | 1 | 7.9 |
| gzip | 3.6 | 2.1 | 3.7 |
| gzip+sort | 2.3 | 3.3 | 2.4 |
| Vertica | 0.6 | 12.5 | 0.6 |
| **Customer Data** | | | |
| Raw CSV | 6200 | 1 | 32.5 |
| gzip | 1050 | 5.9 | 5.5 |
| Vertica | 418 | 14.8 | 2.2 |

Table 4: Compression achieved in Vertica for 1M Random Integers and Customer Data.

- **Bulk Loading and Rejected Records**: Handling input data from the bulk loader that did not conform to the defined schema in a large distributed system turned out to be important and complex to implement.

# 8. PERFORMANCE MEASUREMENTS

## 8.1 C-Store

One of the early concerns of the Vertica investors was that the demands of a product-grade feature set would degrade performance, or that the performance claims of the C-Store prototype would otherwise not generalize to a full commercial database implementation. In fact, there were many features to be added, any of which could have degraded performance such as support for: (1) multiple data types, such as FLOAT and VARCHAR, where C-Store only supported INTEGER, (2) processing SQL NULLs, which often have to be special cased, (3) updating/deleting data, (4) multiple ROS and WOS stores, (5) ACID transactions, query optimization, resource management, and other overheads, and (6) 64-bit instead of 32-bit for integral data types.

Vertica reclaims any performance loss using software engineering methods such as vectorized execution and more sophisticated compression algorithms. Any remaining overhead is amortized across the query, or across all rows in a data block, and turns out to be negligible. Hence, Vertica is roughly twice as fast as C-Store on a single-core machine, as shown in table 3. [8]

## 8.2 Compression

This section describes experiments that show Vertica's storage engine achieves significant compression with both contrived and real customer data. Table 4 summarizes our results which were first presented here [6].

### 8.2.1 1M Random Integers

In this experiment, we took a text file containing a million random integers between 1 and 10 million. The raw data is 7.5 MB because each line is on average 7 digits plus a newline. Applying gzip, the data compresses to about 3.6 MB, because the numbers are made of digits, which are a subset of all byte representations. Sorting the data before applying gzip makes it much more compressible resulting in a compressed size of 2.2 MB. However, by avoiding strings and using a suitable encoding, Vertica stores the same data in 0.6 MB.

---
[8]Comparison on a cluster of modern multicore machines was deemed unfair, as the C-Store prototype is a single-threaded program and cannot take advantage of MPP hardware.

### 8.2.2 200M Customer Records

Vertica has a customer that collects metrics from some meters. There are 4 columns in the schema: **Metric**: There are a few hundred metrics collected. **Meter**: There are a couple of thousand meters. **Collection Time Stamp**: Each meter spits out metrics every 5 minutes, 10 minutes, hour, etc., depending on the metric. **Metric Value**: A 64-bit floating point value.

A baseline file of 200 million comma separated values (CSV) of the meter/metric/time/value rows takes 6200 MB, for 32 bytes per row. Compressing with gzip reduces this to 1050 MB. By sorting the data on metric, meter, and collection time, Vertica not only optimizes common query predicates (which specify the metric or a time range), but exposes great compression opportunities for each column. The total size for all the columns in Vertica is 418MB (slightly over 2 bytes per row). **Metric**: There aren't many. With RLE, it is as if there are only a few hundred rows. Vertica compressed this column to 5 KB. **Meter**: There are quite a few, and there is one record for each meter for each metric. With RLE, Vertica brings this down to a mere 35 MB. **Collection Time Stamp**: The regular collection intervals present a great compression opportunity. Vertica compressed this column to 20 MB. **Metric Value**: Some metrics have trends (like lots of 0 values when nothing happens). Others change gradually with time. Some are much more random, and less compressible. However, Vertica compressed the data to only 363MB.

# 9. RELATED WORK

The contributions of Vertica and C-Store are their unique combination of previously documented design features applied to a specific workload. The related work section in [21] provides a good overview of the research roots of both C-Store and Vertica prior to 2005. Since 2005, several other research projects have been or are being commercialized such as InfoBright [3], Brighthouse [24], Vectorwise [1], and MonetDB/X100 [10]. These systems apply techniques similar to those of Vertica such as column oriented storage, multi-core execution and automatic storage pruning for analytical workloads. The SAP HANA [14] system takes a different approach to analytic workloads and focuses on columnar in-memory storage and tight integration with other business applications. Blink [23] also focuses on in-memory execution as well as being a distributed shared-nothing system. In addition, the success of Vertica and other native column stores has led legacy RDBMS vendors to add columnar storage options [20, 4] to their existing engines.

## 10. CONCLUSIONS

In this paper, we described the system architecture of the Vertica Analytic Database, pointing out where our design differs or extends that of C-Store. We have also shown some quantitative and qualitative advantages afforded by that architecture.

Vertica is positive proof that modern RDBMS systems can continue to present a familiar relational interface yet still achieve the high performance expected from modern analytic systems. This performance is achieved with appropriate architectural choices drawing on the rich database research of the last 30 years.

Vertica would not have been possible except for new innovations from the research community since the last major commercial RBDMs were designed. We emphatically believe that database research is not and should not be about incremental changes to existing paradigms. Rather, the community should focus on transformational and innovative engine designs to support the ever expanding requirements placed on such systems. It is an exciting time to be a database implementer and researcher.

## 11. ACKNOWLEDGMENTS

The Vertica Analytic Database is the product of the hard work of many great engineers. Special thanks to Goetz Graefe, Kanti Mann, Pratibha Rana, Jaimin Dave, Stephen Walkauskas, and Sreenath Bodagala who helped review this paper and contributed many interesting ideas.## 12. REFERENCES

[1] Actian Vectorwise. http://www.actian.com/products/vectorwise.
[2] HP Completes Acquisition of Vertica Systems, Inc. http://www.hp.com/hpinfo/newsroom/press/2011/110322c.html.
[3] Infobright. http://www.infobright.com/.
[4] Oracle Hybrid Columnar Compression on Exadata. http://www.oracle.com/technetwork/middleware/bi-foundation/ehcc-twp-131254.pdf.
[5] PostgreSQL. http://www.postgresql.org/.
[6] Why Verticas Compression is Better. http://www.vertica.com/2010/05/26/why-verticas-compression-is-better.
[7] A. Thusoo, J.S. Sarma, N. Jain, Z. Shao, P. Chakka, S. Anthony, H. Liu, P. Wyckoff and R. Murthy. Hive - A Warehousing Solution Over a MapReduce Framework. *PVLDB*, 2(2):1626–1629, 2009.
[8] D. J. Abadi, D. S. Myers, D. J. Dewitt, and S. R. Madden. Materialization Strategies in a Column-Oriented DBMS. In *ICDE*, pages 466–475, 2007.
[9] B. Chattopadhyay, L. Lin, W. Liu, S. Mittal, P. Aragonda, V. Lychagina, Y. Kwon and M. Wong. Tenzing: A SQL Implementation On The MapReduce framework. *PVLDB*, 4(12):1318–1327, 2011.
[10] P. A. Boncz, M. Zukowski, and N. Nes. MonetDB/X100: Hyper-Pipelining Query Execution. In *CIDR*, pages 225–237, 2005.
[11] S. Ceri and J. Widom. Deriving Production Rules for Incremental View Maintenance. In *VLDB*, pages 577–589, 1991.
[12] J. Dean and S. Ghemawat. MapReduce: Simplified Data Processing on Large Clusters. In *OSDI*, pages 137–150, 2004.
[13] G. DeCandia, D. Hastorun, M. Jampani, G. Kakulapati, A. Lakshman, A. Pilchin, S. Sivasubramanian, P. Vosshall, and W. Vogels. Dynamo: Amazon's Highly Available Key-value Store. In *SOSP*, pages 205–220, 2007.
[14] F. Färber, S. K. Cha, J. Primsch, C. Bornhövd, S. Sigg, and W. Lehner. SAP HANA Database: Data Management for Modern Business Applications. *ACM SIGMOD Record*, 40(4):45–51, 2012.
[15] J. Gray and A. Reuter. *Transaction Processing: Concepts and Techniques*. Morgan Kaufmann Publishers Inc., 1992.
[16] P. J. Haas, J. F. Naughton, S. Seshadri, and L. Stokes. Sampling-Based Estimation of the Number of Distinct Values of an Attribute. In *VLDB*, pages 311–322, 1995.
[17] W. Kim. On Optimizing a SQL-like Nested Query. *ACM TODS*, 7(3):443–469, 1982.
[18] R. Kimball and M. Ross. *The Data Warehouse Toolkit: The Complete Guide to Dimensional Modeling*. Wiley, John & Sons, Inc., 2002.
[19] A. Lakshman and P. Malik. Cassandra: A Decentralized Structured Storage System. *SIGOPS Operating Systems Review*, 44(2):35–40, 2010.
[20] P.-Å. Larson, E. N. Hanson, and S. L. Price. Columnar Storage in SQL Server 2012. *IEEE Data Engineering Bulletin*, 35(1):15–20, 2012.
[21] M. Stonebraker, D. J. Abadi, A. Batkin, X. Chen, M. Cherniack, M. Ferreira, E. Lau, A. Lin, S. Madden and E. J. O'Neil et.al. C-Store: A Column-oriented DBMS. In *VLDB*, pages 553–564, 2005.
[22] G. Moerkotte. Small Materialized Aggregates: A Light Weight Index Structure for data warehousing. In *VLDB*, pages 476–487, 1998.
[23] R. Barber, P. Bendel, M. Czech, O. Draese, F. Ho, N. Hrle, S. Idreos, M.S. Kim, O. Koeth and J.G. Lee et.al. Business Analytics in (a) Blink. *IEEE Data Engineering Bulletin*, 35(1):9–14, 2012.
[24] D. Slezak, J. Wroblewski, V. Eastwood, and P. Synak. Brighthouse: An Analytic Data Warehouse for Ad-hoc Queries. *PVLDB*, 1(2):1337–1345, 2008.
[25] M. Staudt and M. Jarke. Incremental Maintenance of Externally Materialized Views. In *VLDB*, pages 75–86, 1996.
[26] M. Stonebraker. One Size Fits All: An Idea Whose Time has Come and Gone. In *ICDE*, pages 2–11, 2005.
[27] J. D. Ullman. *Principles of Database and Knowledge-Base Systems, Volume II*. Computer Science Press, 1989.